# BREAKING BARRIERS: ASSISTIVE TECHNOLOGY TOOL AS EDUCATIONAL SOFTWARE TO SUPPORT WRITING

Onintra Poobrasert and Waragorn Gestubtim

National Electronics and Computer Technology Center (NECTEC)

**ABSTRACT:**

The preliminary report by Siriraj Hospital suggested that 6% of population who are students in Thailand could be estimated to have learning disabilities (LD). It is therefore necessary for our institute to develop suitable ICT technologies to assist the education of these learning disabilities children. We therefore developed a program called Thai Word Prediction Program. Thai Word Prediction program aims to assist students with learning disabilities in their writing. After the usability engineering, we conducted the experiment with students with learning disabilities at the School in Bangkok. Hence, the results indicated that all three students with learning disabilities in this study improved their ability of writing by 50%, 81.89% and 100% respectively.

## [I] INTRODUCTION

Learning Disabilities or LD which is not caused by physical or intellectual impairment or retardation but rather resulted from various factors adversely affecting the brain e.g. genetic or inheritable factor or impact on brain development during pregnancy or toxic substances which precludes the brain from normally functioning or interpreting data.

Additionally, learning disability can cause a person to have trouble learning and using certain skills. Learning disabilities can be divided into three broad categories (i) developmental speech and language disorders, (ii) academic skills disorders, and (iii) other that include certain coordination disorders and learning handicaps not covered by the other terms [1]. Learning Disabilities vary from person to person however the skills most often affected are reading, writing, listening, speaking, reasoning and calculating. Student with learning disabilities are therefore not able to meet learning achievement although some have high potential as they experience difficulty in the skills as mentioned earlier.

Most researchers ponder that learning disabilities are caused by differences in how a person's brain works and how it processes information [2]. Children with learning disabilities are not *loony* or *lazy*. They truly have average or above average intelligence. Their brains just process information differently. Some learning disabilities student can learn in the same class with normal student but they need certain learning procedure that can elicit their proficiency and talent to offset or cover their disadvantage in order to allow them to live their life normally. Training may help improving reading and calculation skills to some extent but the nature of how such learning disabilities student learn may differ from other student and so technique will be different including technological learning aids such as cassette, video, computer, software program as well as assistive technology [3]. Assistive technology will help increasing ability or adjusting proficiency of learning disabilities student to learn effectively. Moreover, assistive technology can also be applied by the instructor in providing learning opportunity.

Presently, there are many more types of assistive technologies for student with learning disabilities in North America and Europe. However, there is lack of use of assistive technology in educational system in Thailand. Most of students with learning disabilities in Thailand still have to learn under the same methods as those of normal student as there are shortage in Assistive technology or AT suitable for learning. Whilst the number of students with learning disabilities has increasing, our institute deliberates this issue and is committed to developing assistive technology for students with learning disabilities. Therefore the Development of Reading, Writing, Intellectual Capacity, and Calculating Aid Tools for Student with Learning Disabilities Project has begun.

The development process will be divided into three different areas (i) assistive technology to enhance students with learning disabilities in their writing, (ii) assistive technology to enhance students with learning disabilities in their reading, and (iii) assistive technology to enhance students with learning disabilities in their calculating. We, then develop *Thai Word Prediction* program as our first assistive technology program to assist students with learning disabilities in Thailand who struggle with their writing.

## [II] MATERIALS AND METHODS
### 2.1. The Program
There are many types of word predictions such as Aurora Suite, CO: Writer SOLO, Speak Q, Text Help, and Word Q [4] available in the market however most of them are applicable to English speakers. As Thai language is unique therefore we could not use those word prediction programs with students with learning disabilities in Thailand. Not only the language is barrier but the price of the assistive technology is also expensive. Thai Word Prediction program is the solution to solve these problems.

### 2.2. The Conceptual Model
System operation starts when the user wishes to type their work where Thai Word Prediction program is applied in combination with any word processor on Windows. Thai Word Prediction program will run as a background of word processor but program results will be displayed in Top Level, that is, program results will be displayed at top level of Windows which can be monitored at all time. When the user hits the keys for typing, the system interface part will function by detecting signal from such keystroke in order to forward such signal for processing and selecting vocabulary from database. Vocabulary searching will be done by considering frequency of words which have been sort out as ranking and are then partially selected as set out to display to the user for selection. Upon such display by the program of a list of words for selection by the user and the user has selected certain vocabulary from such list, system interface will send such vocabulary to word processor for further printing immediately.

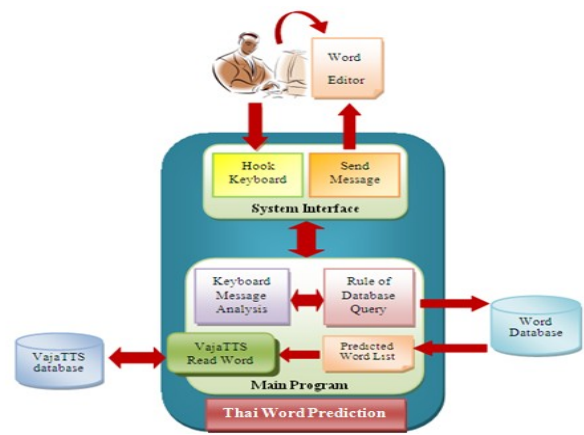

Fig. 1. The conceptual model of Thai word prediction.

At the same time, next list of words will be shown to the user for further selection where such selected vocabulary will be given additional frequency value in the database. There are 2 types of word lists that the program offers to the students: *word completion* and *word prediction*. In the event that there is no such vocabulary in the database, the user can add such vocabulary by hitting the function key designated for adding vocabulary. In addition, the advantage of Thai Word Prediction program is that it can pronounce vocabulary in a list of vocabulary. The overview of the conceptual model of Thai Word Prediction program is as demonstrated in Figure 1.

### 2.3. The Program Design

Program design can be separated into five components as follows:

*Vocabulary database component:* It is divided into general vocabulary database and Thai lesson vocabulary database. In respect of design of vocabulary database of Thai Word Prediction program, its objective is to obtain small size program. Therefore, vocabulary data will be systematically stored in the pattern of .txt files instead of Paradox.

*Keystroke data interception component:* It can help the system to intercept message from keystroke for the purpose of processing value from such keystroke.

*Component of sending message for printing by word processor currently applied by the user (Send Message)*: Its function is to send vocabulary selected from word list by the user to be printed by word processor currently applied by the user.

*Core program component:* The core program will operate by processing data in order to search for list of possible words when retrieving data (vocabulary) from Hook keyboard. Vocabulary will be explored from database based on the frequency of words and then put in order from most to least frequent and after that the system will select part of them to display as list of possible words for the user.

*Pronunciation component:* Thai Word Prediction program adopts VAJA TTS software used for synthesizing Thai speaking voice in API. VAJA TTS developed by Speech and Audio Technology Lab (SPT) [5], National Electronics and Computer Technology Center (NECTEC).

### 2.4. The User Interface Design

With respect to program pattern, when it is opened, a window will show up as seen in Figure 2. This window requires the user to fill in username and choose type of vocabulary databases which are BEST Corpus (Best corpus consists of 4 categories such as Article, Encyclopedia, News, and Novel) and Corpus from Thai lesson vocabulary (secondary level). Another important part used in the program is the component of Thai Word Prediction. The component of Thai Word Prediction consists of four parts; menu button, keyboard parameter display, list of words and button for switching page of list of words (see Figure 3).

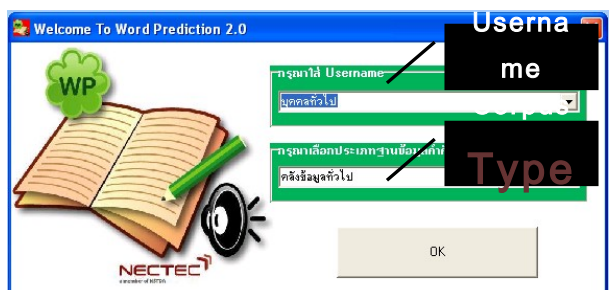

**Fig. 2. Username and corpus box.**

Manuscript received on 28th November 2013, revised on 12th December 2013
**(Onintra Poobrasert and Waragorn Gestubtim)**

Fig. 3. The Component of Thai word prediction.

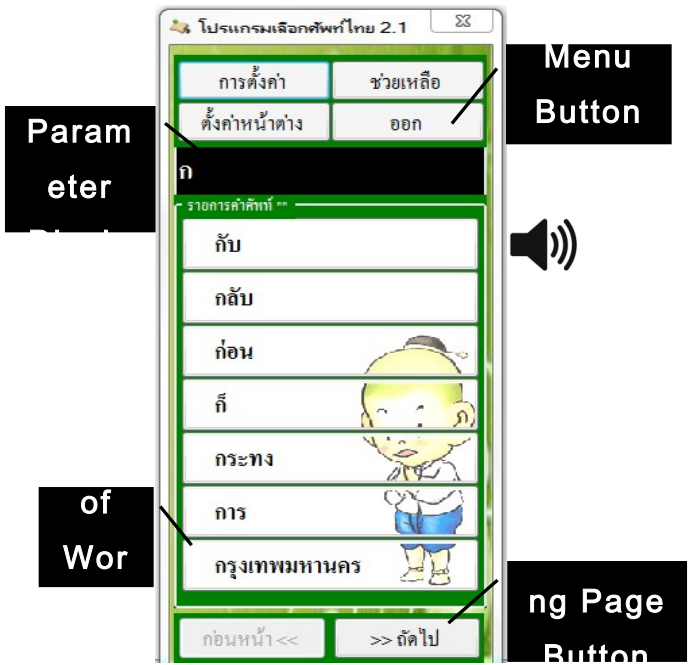

## 2.5. The Usability Testing

The usability testing was conducted and questionnaires were developed and given to four information technology students who had taken computer science, information system and information technology courses from the International Institute of Technology in Thailand.

According to Nielsen [6], the best usability testing results come from testing no more than 5 users [$U = 1 − (1 − p)^n$, where p is the probability of one subject identifying a specific problem and n is the number of subjects or test sessions]. We developed a questionnaire that would enable computer users to perform heuristic evaluation. Heuristic evaluations usually are conducted by a small set (one to three) of evaluators [7]. The evaluators independently examine a user interface and judge its compliance with a set of usability principles which based on Nielsen's usability testing [8]. In this study we applied Nielsen's usability inspection [9] and Shneiderman's principles for designing the user interface [10] to the design and development of Thai Word Prediction.

Fig. 4. Ease of use of program.

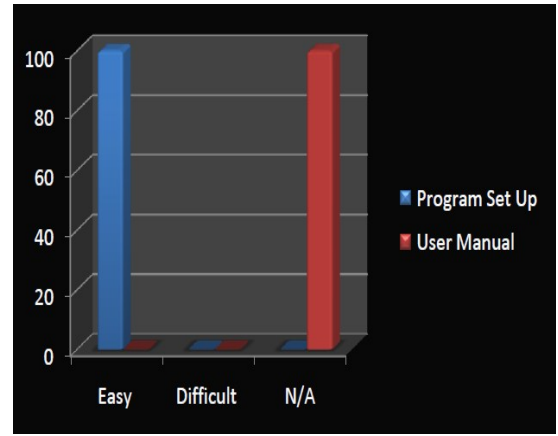

Usability testing was divided into 3 scenarios, each of which will evaluate a different aspect of the testing. The first aspect was ease of use of program. The second aspect was technical aspects of the program and the last aspect was management of the program [11,12].

*Ease of use of program:* all examiners are of the opinion that there have none of the information for user manual for the program. In this regard the developer is working out for such concern. In addition, most of the examiners view that the user can independently use the program and installation process is not complicated (see Figure 4).

*Technical aspects of the program:* all examiners agree that it is appropriate to set out on-off function for acoustic and also agree for changing location of vocabulary list as the user wishes. All examiners agree for ability to open many files and to run word guessing program at the same time. In addition, all examiners agree with ability to quit program anytime during use. Most of the examiners do not agree with that program can change font size, color and type. In this regard, the developer wishes the program to be able to change font size, color or type in accordance with the presetting theme selected by the user. Most of the examiners do not agree with ability to change voice speed and also view that program still has errors/bugs (see Figure 5).

*Program Management:* all examiners are in the agreement for accuracy of word guessing, automatic word completion and addition of vocabulary in the list (see Figure 6).

**Fig. 5. Technical aspects of the program.**

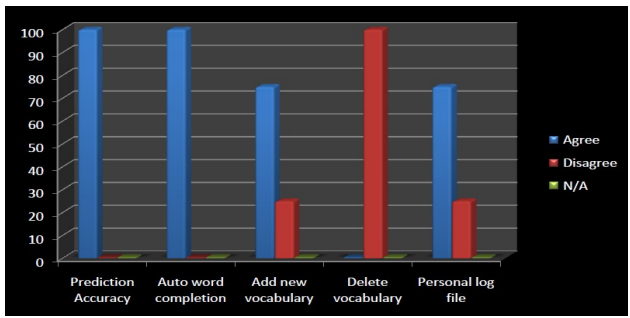

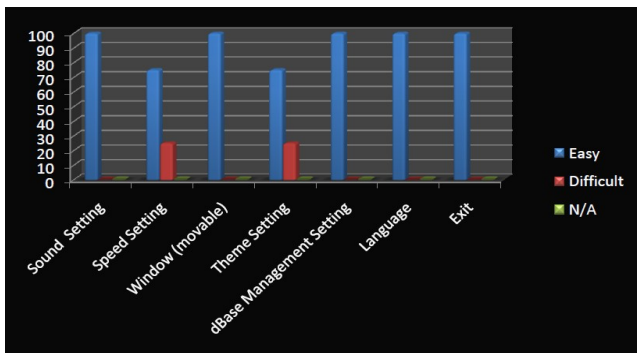

**Fig. 6. Program management.**

## 2.6. Comments on Usability Testing

After the usability testing of Thai Word Prediction, problems found on the application of such program and their solution can be summarized as follows (i) Application of *database* of the program: although it is easy to develop program, there are many subsequent problems. For instance, it is required to have large memory and it is restricted by limitation of size of database. In addition, a problem is also found when program is installed in the machine with different OS. In order to solve these problems, it is necessary to design new pattern for vocabulary searching under the condition that all problems resulted from the application of *database* must be solved and processing speed is to be better than or similar to the application of database, (ii) some parts of GUI pattern of the program are still not suitable for the user or not completed, and (iii) modules of Thai Word Prediction should designed as API in order to make it easy for further development in the future.

We, then solve all the problems mentioned above before conducting the experiment with real users (students with learning disabilities).

**Fig. 7. The comparison number of errors between pretest and posttest of essay writing.**

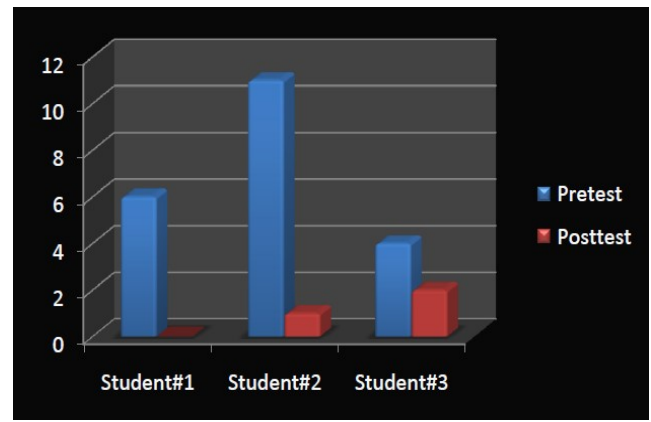

## [III] EXPERIMENT

Thai Word Prediction program was experimentally applied to five students with learning disabilities who struggle with writing in grade 5th at the school in Bangkok.

### 3.1. Method for selecting children
- Students with learning difficulties who struggle with writing in grade 5$^{th}$ were selected to conduct Pretest number one (There were 25 students in total).
- Students' writing ability was tested on the subject of Father's Day where each of which was allocated about 20 minutes to finish the test.
- Writing test was designated to be an essay where students had been taught about the pattern of essay from Thai instructor and the teacher explained to them again that it should have three parts; introduction, content and summary.
- Researcher collected writing results from the students and then selected ten out of them.
- Thai instructor gave advice on selection of students to finally get five students which were a suitable number as this test was done on individual basis and thus large group of student was not needed (adequate number of equipment for children under test, manageable by instructors).

### 3.2. Experiment method
- Researcher installed the program.
- Researcher tested computer skills of such five selected students.
- Researcher explained principle and rationale, background and objective of the project.
- Researcher explained the application of Thai Word Prediction program.
- Researcher explained how to practice and distribute typing practice schedule and file saving.
- Researcher started Pretest number two by assigning ten vocabularies for typing.
- Researcher assigned the instructors to control practice and then came in to follow up the next three days.

### 3.3. Practicing method
- Student typed the message read to him/her by researcher, instructor or other student.
- Type 1$^{st}$ message on Notepad without opening Thai Word Prediction program. (Save the file as 1.1no_wp.txt).
- Type 1$^{st}$ message on Notepad while opening Thai Word Prediction program. (Save the file as 1.1w_wp.txt).

The total time for practice included 30 sessions with two different messages alternately used for typing practice.

## [IV] RESULTS AND DISCUSSION

Table 1 show the summary of results in the comparison of keystrokes from typing practice within ten sessions of the two messages by Thai Word Prediction Program.

From Table 1, analysis of results can be summarized as follows:
- Thai Word Prediction Program could help Student #1 to reduce keystroke while typing 1$^{st}$ message at the average of 49.31% and keystroke while typing 2$^{nd}$ message at the average of 11.83%. From the observation, Student #1 could mostly choose correct words from list of vocabulary. To do so, Student #1 based her selection on pronunciation made by the program while some words were selected from list of word as she could remember their spelling when she saw such words. Student #1 is observant and able to learn fast with excellent memory.

*Student #1 has IQ level slightly higher than the average.

*Student #1 suffers from the condition of Attention Deficit Hyperactivity Disorders (ADHD).

- Thai Word Prediction Program could help Student #2 to reduce keystroke while typing 1$^{st}$ message at the average of 12.87% and keystroke while typing 2$^{nd}$ message at the average of 25.51%. Student #2 typing skill was in excellent level which sometime allowed him to disregard vocabulary choosing assistive technology during his typing. However, there were many typing errors during the practice and researcher had to urge him to wait and see list of vocabulary. In general, if Student #2 allowed the program to help him, he could correctly choose vocabularies after listening to pronunciation.

*Student #2 has IQ level higher than the average (High Average Intelligence).

*Student #2 suffers from the condition of Attention Deficit Hyperactivity Disorders (ADHD: slight haste).

- Thai Word Prediction Program could help Student #3 to reduce keystroke while typing 1$^{st}$ message at the average of 22.95% and keystroke during typing 2$^{nd}$ message at the average of 18.23%. Student #3 had moderate typing skills and if he was encouraged, he would try to use the program during his typing. There were many typing errors if Student #3 failed to use the program but he tried to practice it. Researcher was told by the instructor that Student #3 showed better development.

The comparison number of errors between Pretest and Posttest of essay writing by the students represent in Figure 7. From Figure 7, analysis of results can be concluded as follows:

- From essay writing by Student #1 (Pretest) without the program, there were 6 errors while her typing with the program (Posttest) showed no error in her essay writing. Therefore, Student #1 had development in correctly typing and vocabulary choosing at 100% level which was deemed as an excellent improvement.

- From essay writing by Student # 2 (Pretest) without the program, there were 11 errors while his typing with the program (Posttest) showed almost completely correct in his essay writing with only one error. Therefore, Student #2 had development in correctly typing and vocabulary choosing at 81.89% which was deemed as good improvement.

- From essay writing by Student #3 (Pretest) without the program, there were 4 errors while his typing with the program (Posttest) showed only two errors in his essay writing. Therefore, Student #3 had development in correctly typing and vocabulary choosing at 50% level. However, Student #3 had limited ability to write essay and thus this conclusion was based on the essay typed by him at that time.

**[Table-1]**.

| Name | Message#1: Total Keystrokes Used | Keystrokes Reduce | Reduction (%) | Messaget#2: Total Keystrokes Used | Keystrokes Reduce | Reduction (%) |
|---|---|---|---|---|---|---|
| Student #1 | 219 | 108 | 49.31 | 277.2 | 32.8 | 11.83 |
| Student #2 | 289.7 | 37.3 | 12.87 | 247 | 63 | 25.51 |
| Student #3 | 265.8 | 61.2 | 22.95 | 262.2 | 47.8 | 18.23 |

Table: 1. Comparison of keystrokes from typing practice of the two messages.

## [V] CONCLUDING REMARK

In conclusion, most users were in agreement with the advantages from the assistive technology; Thai Word Prediction which could help select words well and fast. It also assisted the users to be able to choose vocabulary and print the work correctly. Most users agreed that the program could help the children to prepare their report faster and to pronounce word better. For some students whose typing skill was fair, it took time to look for alphabet on keyboard but this program could make it faster by just typing some alphabets and then it can make a guess by listing the word for the user. This could significantly help students with learning disabilities save time for their typing and select exact word. In addition, the students with learning disabilities agreed that Thai Word Prediction was simple, not complicated and convenient for them during typing work.

Moreover, most parents of the students with learning disabilities were in the agreement that the program could also help their children to know how each word is spelled as it provides pronunciation for each word. At the same time some parents were of the view that pronunciation by the program for some word was deviated but most of them admitted that the program could help their children to be able to pronounce vocabulary.

## [VI] FUTURE WORK

In the future, Thai Word Prediction feature can be applied to work in combination with other programs e.g. Thai Spell Checker, Thai Word Spelling, and Thai Word Processor in order to enhance students with learning disabilities in their learning as well as to optimize their accurate writing.

### FINANCIAL DISCLOSURE

This project was funded by the Cluster and Program Management Office (CPMO), National Science and Technology Development Agency (NSTDA) P-00-40230.

### ACKNOWLEDGEMENT

We would like to convey our thanks and acknowledge the assistance of Speech and Audio Technology Lab (SPT), National Electronics and Computer Technology Center (NECTEC), Miss Wantanee Phantachat, Dr. Putthachart Pothibal, and Dr. Somporn Warnset. Our thanks extend to the Director, the teachers and the students at the School in Bangkok for their time and assistance.

### REFERENCES

[1] Vogel, A., S., and Reder, M. (1998). *Learning Disabilities, Literacy and Adult Education*. Brookes, 1st Ed. ISBN 978-1557663474.

[2] Bradley, R., Danielson, L., Hallahan P.,D. (2013). *Identification of Learning Disabilities: Research to Practice*. Routledge. ISBN 9781135627638.

[3] Richards, G. (1999). *The RET Assessment for Dyslexia* in *The Source for Dyslexia and Dysgraphia*. Linguisystems, Inc., East Moline, Illinois.

[4] Stanberry, K., Raskind., M. (2010). *Word prediction software programs: Learn about assistive technology tools called word prediction software programs*. Retrieved [January 2013] from the World Wide Web: http://www.greatschools.org/specialeducation/assistive-technology/966word-prediction-software-programs.gs?page=1.

[5] Speech and Audio Technology Lab (SPT), National Electronics and computer Technology Center (NECTEC). Retrieved [August 2013] from the World Wide Web: http://vaja.nectec.or.th/

[6] Nielsen, J. (1994). *Heuristic Evaluation*. In Nielsen, J., and Mack, R.L. (Eds.), Usability Inspection Methods, John Wiley & Sons, New York, NY.

[7] Rosson, M. & Carroll, J. (2010). *Scenario-Based Usability Engineering* (Paperback). Morgan & Claypool Publishers. San Rafael, CA.

[8] Nielsen, J., and Molich, R. (1990b). Heuristic evaluation of user interfaces, *Proc. ACM CHI'90 Conf.* (Seattle, WA, 1-5 April), 249-256.

[9] Nielsen, J. (1995). *Usability Engineering*: AP Professional. Chestnut Hill: Academic Press.

[10] Shneiderman, B., Jacobs, S., Plaisant, C., & Cohen, M. (2009). *Designing the User Interface: Strategies for Effective Human-Computer Interaction* 5[th] edition: N.J: Addison Wesley.

[11] Poobrasert, O. (2010). Employing Usability Engineering in the Development of Assistive Technology Tools for People with Learning Disabilities. In *Proceedings of World Conference on Educational Multimedia, Hypermedia and Telecommunications*, Chesapeake, VA: AACE.

[12] Poobrasert, et al. (2011). Technology-enhanced Learning for Students with Learning Disabilities. Retrieved [August 2013] From the World Wide Web: http://doi.ieeecomputersociety.org/10.1109/ICALT.2011.154.